\documentclass{PoS}

\usepackage{amsmath,amsthm}

\title{Measurement of the Extragalactic Background Light with VERITAS}

\ShortTitle{Measurement of the Extragalactic Background Light with VERITAS}

\author{\speaker{Elisa Pueschel}, for the VERITAS Collaboration\footnote{https://veritas.sao.arizona.edu/}
\footnote{for collaboration list see PoS(ICRC2019)1177}\\
        Deutsches Elektronen-Synchrotron (DESY) \\
        Platanenalle 6, D-15738 Zeuthen, Germany\\
        E-mail: \email{elisa.pueschel@desy.de}}

\abstract{The extragalactic background light records the history of infrared, optical and ultraviolet light radiation including re-radiation since the epoch of reionization. While challenging to measure directly, it can be measured indirectly via its impact on observed spectra of extragalactic gamma-ray emitters. VERITAS, a ground-based imaging atmospheric-Cherenkov telescope array sensitive to gamma rays above 100 GeV, has accrued 10 years of observations of hard-spectrum blazars. The energy and redshift range covered enables the measurement of the EBL in the range 0.56-56~$\mu$m, accessing the poorly constrained cosmic infrared background region. New constraints on the EBL resulting from the joint analysis using 16 spectra from 14 VERITAS-observed blazars will be presented. The method is independent of assumptions about the shape of the EBL spectrum, and includes a full treatment of systematic and statistical uncertainties. The measured spectrum is in good agreement with lower limits from galaxy counts, limiting the potential contribution from a diffuse component.}

\FullConference{36th International Cosmic Ray Conference -ICRC2019-\\
		July 24th - August 1st, 2019\\
		Madison, WI, U.S.A.}

\begin{document}

\section{Introduction}
The Universe beyond redshift $z\sim$1 is opaque to very-high-energy ($>$100~GeV) gamma rays, as gamma rays undergo pair production on the extragalactic background light (EBL) en route to the observer. This process attenuates the flux of gamma rays reaching the observer, with attenuation increasing with energy and redshift. This complicates the study of the acceleration and emission mechanisms of extragalactic gamma-ray emitters like blazars, as the intrinsic energy spectra are inaccessible. A more precise measurement of the EBL would reduce one source of uncertainty in such studies, and would also be relevant from the perspective of cosmology, as the EBL tracks the production and re-radiation of light in the infrared to ultraviolet range since the onset of galaxy formation~\cite{Krennrich2013}.

Direct measurements of the EBL spectral energy distribution (SED) are commonly treated as upper limits, as foreground contamination can impact the measurements (\textit{e.g.} \cite{CIBER, NewHorizons}). Attempts to calculate the EBL SED based on adding up the light expected based on galaxy surveys are taken as lower limits (\textit{e.g.} \cite{Driver2016}), due to their inability to account for additional components, for instance from unresolved source populations or dark matter decay. State of the art theoretical models, such as those described in Refs.~\cite{Franceschini2008, Dominguez2011, Gilmore2012}, are in reasonable agreement with the lower limits from galaxy counts. 

Constraining or measuring the EBL indirectly from observations of extragalactic gamma-ray sources evades the limitations of the approaches mentioned above, but in turn is limited by ignorance of the gamma-ray spectra prior to EBL absorption, and hence relies on assumptions about the intrinsic spectra. Such studies typically use high-frequency-peaked BL Lacertae objects (HBLs), due to their detection at a range of redshifts and their gamma-ray emission over a wide energy range, as well as intermediate-frequency-peaked BL Lacertae objects (IBLs) and flat spectrum radio quasars (FSRQs).

\section{VERITAS observations}
The imaging atmospheric Cherenkov technique reconstructs gamma-ray energies and arrival directions via detection of Cherenkov light, which traces extensive air showers in the atmosphere. The VERITAS instrument uses this technique to detect gamma-ray emitters and to reconstruct their morphology and photon spectra. VERITAS is located at the Fred Lawrence Whipple Observatory (FLWO) in southern Arizona (31 40N, 110 57W,  1.3km a.s.l.), and consists of four telescopes of 12~m diameter~\cite{VERITASinstrument}. Each telescope is equipped with a photomultiplier tube camera with 499 pixels. The sensitive energy range of the instrument is $\sim$100 GeV to $>$30 TeV. The energy and angular resolution of VERITAS are comparable to other instruments of its type, with energy resolution of 15-25\% (depending on energy) and a 68\% containment of 0.1$^\circ$  at 1 TeV~\cite{VERITASspecs}.  

VERITAS has collected several hundred hours of observations of hard-spectrum blazars, enabling precise measurements of their photon spectra. Most of the objects included in this study are HBLs, although one IBL object is also included, due to its high redshift and well-measured spectrum. The observation of the objects over multiple seasons makes possible the study of flux variability and its impact on the photon spectra. Table~\ref{sourcelist} lists the blazars included in this analysis.

The sources were analyzed using standard VERITAS calibration and reconstruction pipelines~\cite{GernotICRC2017, BDTpaper}. The detection significances per source are listed in Table~\ref{sourcelist}. All sources other than 1ES 0414+009 displayed flux variability, but only two sources (1ES 1959+650 and 3C 66A) showed a significant difference in the spectral index or energy threshold between flux states. In these cases, the observations were divided to produce photon spectra corresponding to high and low flux states. For the other sources, time-averaged spectra were produced utilizing the full datasets.

The photon spectra were fit assuming three possible spectral shapes: power law, power law with exponential cutoff, and log parabola. For spectra that were well-described by a power law (fit probability of $>$5\%), the more complex models were not considered. Table~\ref{sourcelist} shows the preferred fit model for each of the sources. When two fit models are listed, the first refers to the high flux state while the second refers to the low flux state.

\begin{table}
\centerline{
\begin{tabular}{cccccc}
Target & Redshift & Exposure [min] & $\sigma_{\textrm{detect}}$ & Flux/spectral variability & Fit model \\
\hline
\hline
1ES 2344+514  & 0.044 & 4514 & 31.0 & yes/no & LP \\
1ES 1959+650 & 0.048 & 3324 & 102.5 & yes/yes & PLEC/LP \\
RGB J0710+591 & 0.125 & 7926 & 11.5 & yes/no & PL \\ 
H 1426+428 & 0.129 & 5477 & 13.3 & yes/no & LP \\
1ES 1215+303 & 0.13 & 10071 & 33.0  & yes/no & PL \\
1ES 0229+200 & 0.14 & 8392 & 12.3 & yes/no & PL \\
1ES 1218+304 & 0.182 & 9524 & 63.0 & yes/no & PL \\
1ES 1011+496 & 0.212  & 2160 & 43.5 & yes/no & PL \\
MS 1221.8+2452 & 0.218 & 152 & 22.0 & yes/no & PL \\
1ES 0414+009 & 0.287 & 6457 & 9.3 & no/no & PL \\
1ES 0502+675 & 0.341 & 1970 & 13.9 & yes/no & PL \\
3C 66A & 0.34--0.41 & 5926 & 26.4  & yes/yes & LP/LP \\
PG 1553+113 & 0.43--0.58 & 7774 & 71.8 & yes/no & LP \\
PKS 1424+240 & 0.604  & 10697 & 28.3 & yes/no & LP \\
 \hline
\end{tabular}
}
\caption{\small{The sources included in the analysis are listed, with their redshift, observation time, detection significance, whether the source displayed flux and spectral variability, and the model used to describe the photon spectrum. The abbreviations in the final column denote power law (PL), log parabola (LP) and power law with exponential cutoff (PLEC). When two fit models are listed, the first shows the high flux state while the second shows the low flux state.}}
\label{sourcelist}
\end{table}

\section{EBL analysis}
A procedure of correcting observed spectra with generic EBL SEDs and down-weighting shapes producing unphysical corrected spectra was used to reconstruct the EBL SED. The generic EBL SEDs were randomly generated by considering 12 grid points in EBL wavelength between 0.1 and 100~$\mu$m, and selecting from a flat intensity distribution between 1.0 and 50~nW~m$^{-2}$~sr$^{-1}$ at each grid point. Restrictions were placed on the maximum sharpness of the transitions between grid points; this biased the initially flat distributions, making it necessary to recover the unbiased distributions through the application of weights. A total of 480 000 shapes were generated by drawing second order splines through the EBL wavelength/intensity grid point combinations. 

The EBL opacity over the observed range of energy and redshift was calculated for each EBL shape. The evolution of the EBL with redshift was accounted for in the calculation with an empirical factor, $f_{\textrm{evo}}$, scaling the EBL number density by (1 + $z$)$^{3 - f_{\textrm{evo}}}$ with $f_{\textrm{evo}}$=1.7. This value reproduced the evolution of several theoretical models for the redshifts considered~\cite{Franceschini2008, Dominguez2011, Gilmore2012}.

For each generic EBL shape, the observed spectrum is corrected for absorption to produce an ``intrinsic" spectrum. The corrected spectrum is fit with the same fit models considered for observed spectra, and with the same criterion: if a power law fit gives an acceptable description, further models are not considered. Several requirements are imposed to limit the intrinsic spectra to physically-motivated shapes: the intrinsic spectra cannot become harder with increasing energy (convex shapes), and the intrinsic spectral index $\Gamma$ cannot be too hard ($\Gamma \leq$1) for power law and power law with exponential cutoff models. The fit $\chi^{2}$ is translated into a weight on the associated EBL shape of exp(-$\chi^{2}$/2). The large $\chi^{2}$ associated with poorly fit, unphysical intrinsic spectra thus down-weights the EBL shape used to produce it. The EBL intensity distribution at each wavelength after weighting can be interpreted as a probability density.

To combine the results from all sources observed, the $\chi^{2}$ values are summed and a joint weight applied. The EBL wavelength range over which each source contributes is determined from the energy of the lowest and highest spectral points. The 68\% and 95\% containment bands are calculated for each distribution.

\section{Systematic uncertainties}
Several systematic uncertainties limit the measurement precision, related to the instrument response, the redshift evolution of the EBL, and the knowledge of the source distances.

\noindent 1) \textit{VERITAS energy scale uncertainty and finite energy resolution} 

The VERITAS energy scale uncertainty of 20\%, when combined with the instrument's energy-dependent energy resolution, introduces an uncertainty on reconstructed spectral indices of $\sim$10\%. This is accounted for by varying the limits on the intrinsic spectral index by $\pm$10\% of the observed spectral index and taking the most conservative extracted EBL intensity ranges. This broadens the containment band on the EBL intensity by $\leq$10\%.

\noindent 2) \textit{Treatment of EBL redshift evolution} 

The evolution of the EBL intensity predicted by theoretical models varies. The redshift evolution factor $f_{\textrm{evo}}$ was selected to approximately reproduce the evolution of the models, but the difference between them can be accounted for as an additional uncertainty on the EBL opacity. This difference was calculated for two widely-used EBL models~\cite{Franceschini2008, Gilmore2012}. The uncertainty on the opacity was propagated to an uncertainty on EBL intensity range, broadening the containment band on the EBL intensity by $\leq$12\%.

\noindent 3) \textit{Lack of precise redshift estimate for PG 1553+113 and 3C 66A} 

This uncertainty is accounted for by setting the assumed redshifts to their upper and lower bounds, shown in Table 1, and taking the most conservative extracted EBL intensity ranges.

\section{Results and Discussion}
\begin{figure}
\centerline{\includegraphics[width=0.75\textwidth]{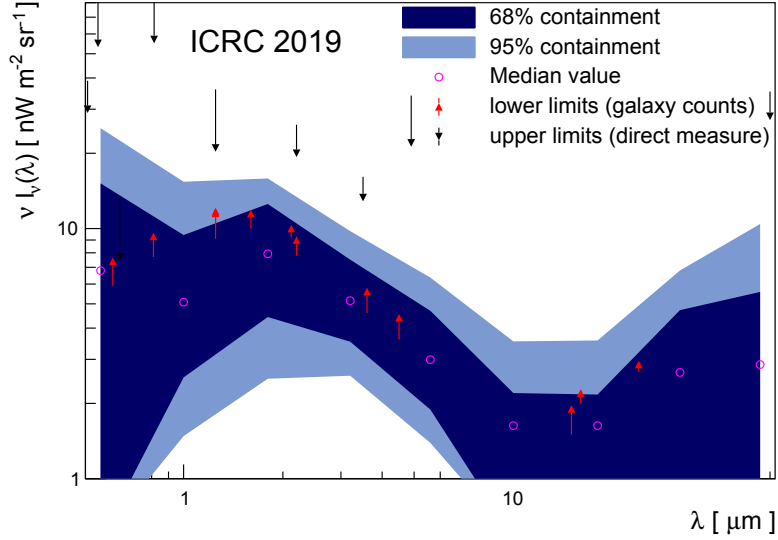}}
\caption{\small{The 68\% and 95\% containment bands on the EBL intensity as a function of wavelength. Also shown are upper and lower limits from direct measurements and galaxy counts, respectively.}}\label{fig:EBLsed}
\end{figure}

Figure~\ref{fig:EBLsed} shows the 68\% and 95\% containment bands on the EBL intensity as a function of wavelength, after inclusion of systematic uncertainties. Also shown are the median values of the probability distributions and the lower limits from galaxy counts (upward-facing arrows) and upper limits from direct measurements (downward-facing arrows). The 68\% containment band deviates little from the galaxy counts measurements, indicating a limited budget for additional components. 

\begin{figure}
\centerline{\includegraphics[width=0.75\textwidth]{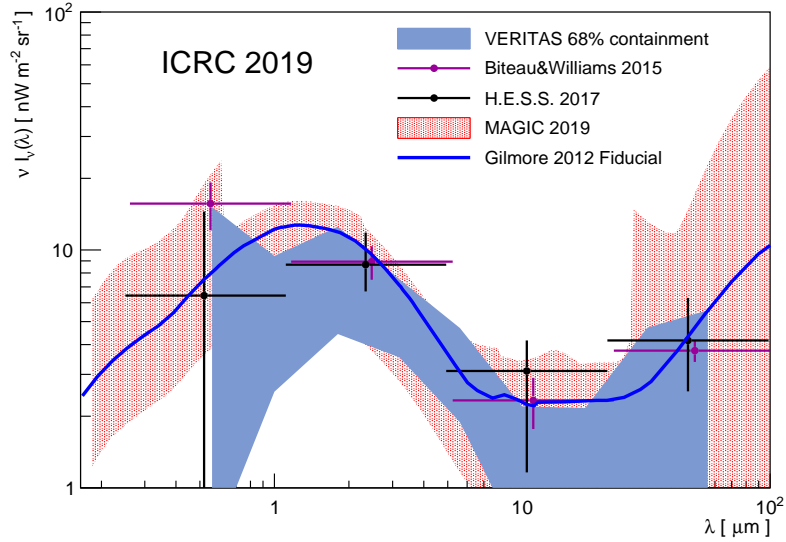}}
\caption{\small{Comparison of the VERITAS 68\% containment band on the EBL intensity with recent measurements from H.E.S.S., MAGIC, and an archival study~\cite{HESS2017, MAGIC2019, Biteau2015}. The theoretical model of Ref.~\cite{Gilmore2012} is shown as a solid line. Figure adapted from Ref.~\cite{VTSEBL}.}}\label{fig:EBLsedcomparison}
\end{figure}

Figure~\ref{fig:EBLsedcomparison} compares the VERITAS results against other recent measurements, from H.E.S.S., MAGIC, and an archival study based on a large ensemble of published spectra~\cite{HESS2017, MAGIC2019, Biteau2015, VTSEBL}. All results shown for comparison show 1$\sigma$ uncertainties and include systematic uncertainties. The model of Ref.~\cite{Gilmore2012} is plotted as an example of a widely-used theoretical model. The VERITAS band is in good agreement with other measurements, and around $\lambda_{\textrm{EBL}}$ of 10$\mu$m gives the strongest constraints. The agreement with the model of Ref.~\cite{Gilmore2012} is also good. On the other hand, the VERITAS results cover a comparatively narrow range in $\lambda_{\textrm{EBL}}$, which could potentially be extended with the inclusion of observations of nearby blazars, such as Mrk 501, Mrk 421, and M87. Taken together, these measurements point to an EBL SED that is well described by the lower limits from galaxy counts, although the uncertainties remain too large to rule out a diffuse component. 

\section*{Acknowledgments} 
This research is supported by grants from the U.S. Department of Energy Office of Science, the U.S. National Science Foundation and the Smithsonian Institution, and by NSERC in Canada. This research used resources provided by the Open Science Grid, which is supported by the National Science Foundation and the U.S. Department of Energy's Office of Science, and resources of the National Energy Research Scientific Computing Center (NERSC), a U.S. Department of Energy Office of Science User Facility operated under Contract No. DE-AC02-05CH11231. We acknowledge the excellent work of the technical support staff at the Fred Lawrence Whipple Observatory and at the collaborating institutions in the construction and operation of the instrument. E. Pueschel acknowledges the support of the Young Investigators Program of the Helmholtz Association.

\end{document}